\documentclass[modern]{aastex62}

\hypersetup{citecolor=blue, filecolor=blue, urlcolor=blue}

\usepackage{braket}
\usepackage{amsmath}
\usepackage{bm}

\newcommand{\acronym}[1]{{\small{#1}}}
\newcommand{\project}[1]{\textsl{#1}}
\newcommand{\apogee}{\project{\acronym{APOGEE}}}
\newcommand{\gaia}{\project{Gaia}}
\newcommand{\wise}{\project{\acronym{WISE}}}
\newcommand{\zmass}{\project{\acronym{2MASS}}}

\revised{\today}

\shorttitle{Dynamical Spiral Perturbation in the Galactic Disk}
\shortauthors{A.-C. Eilers et al.}

\begin{document}\frenchspacing

\title{\textbf{The Strength of the Dynamical Spiral Perturbation in the Galactic Disk}}

\author[0000-0003-2895-6218]{Anna-Christina Eilers}\thanks{NASA Hubble Fellow}
\affiliation{Max Planck Institute for Astronomy, K\"onigstuhl 17, 69117 Heidelberg, Germany}
\affiliation{MIT Kavli Institute for Astrophysics and Space Research, 77 Massachusetts Ave., Cambridge, MA 02139, USA}

\author[0000-0003-2866-9403]{David W. Hogg}
\affiliation{Center for Cosmology and Particle Physics, Department of Physics, New York University, 726 Broadway, New York, NY 10003, USA}
\affiliation{Center for Data Science, New York University, 60 Fifth Ave, New York, NY 10011}
\affiliation{Max Planck Institute for Astronomy, K\"onigstuhl 17, 69117 Heidelberg, Germany}
\affiliation{Center for Computational Astrophysics, Flatiron Institute, 162 5th Avenue, New York, NY, USA}

\author[0000-0003-4996-9069]{Hans-Walter Rix}
\affiliation{Max Planck Institute for Astronomy, K\"onigstuhl 17, 69117 Heidelberg, Germany}

\author[0000-0002-6411-8695]{Neige Frankel}
\affiliation{Max Planck Institute for Astronomy, K\"onigstuhl 17, 69117 Heidelberg, Germany}

\author[0000-0001-8917-1532]{Jason A. S. Hunt}
\affiliation{Center for Computational Astrophysics, Flatiron Institute, 162 5th Avenue, New York, NY, USA}

\author[0000-0002-0030-371X]{Jean-Baptiste Fouvry}
\affiliation{Sorbonne Université, CNRS, UMR 7095, Institut d’Astrophysique de Paris, 98 bis Bd. Arago, 75014 Paris, France}

\author[0000-0003-2027-399X]{Tobias Buck}
\affiliation{Leibniz-Institut f\"ur Astrophysik Potsdam (AIP), An der Sternwarte 16, 14482 Potsdam, Germany}

\correspondingauthor{Anna-Christina Eilers}
\email{eilers@mit.edu}

\begin{abstract}
The mean Galactocentric radial velocities $\langle v_{R}\rangle(R,\varphi)$ of luminous red giant stars within the mid-plane of the Milky Way reveal a spiral signature, which could plausibly reflect the response to a non-axisymmetric perturbation of the gravitational potential in the Galactic disk. We apply a simple steady-state toy model of a logarithmic spiral to interpret these observations, and find a good qualitative and quantitative match. Presuming that the amplitude of the gravitational potential perturbation is proportionate to that in the disk's surface mass density, we estimate the surface mass density amplitude to be $\Sigma_{\rm max} (R_{\odot})\approx 5.5\,\rm M_{\odot}\,pc^{-2}$ at the solar radius when choosing a fixed pattern speed of $\Omega_{\mathrm p}=12\,\rm km\,s^{-1}\,kpc^{-1}$. Combined with the local disk density, this implies a surface mass density contrast between the arm and inter-arm regions of approximately $\pm 10\%$ at the solar radius, with an increases towards larger radii. Our model constrains the pitch angle of the {\it dynamical} spiral arms to be approximately $12^{\circ}$. 
\end{abstract}

\keywords{Galaxy: disk, structure, kinematics and dynamics -- stars: distances}

\section{Introduction}

Many external disk galaxies in the universe show a spiral structure in the emission of stars and gas \citep[e.g.][]{Hubble1936, GalaxyZoo2011, Conselice2014}. The spiral arms constitute the sites of star formation and thus comprise many luminous young stars. Evidence for a spiral pattern within our own Galaxy, the Milky Way, dates back to the 1950s and is based on distance measurements towards OB star associations \citep{Morgan1953}, as well as $21$~cm observations from Galactic atomic hydrogen (\ion{H}{1}) \citep{Oort1952, VandeHulst1954}. Since then there have been numerous attempts to determine the precise locations and densities of the spiral arms by means of molecular masers associated with massive young stars \citep{Reid2014, Reid2019}, \ion{H}{1} regions \citep{Levine2006}, photometric star counts \citep[e.g.][]{Binney1997, Benjamin2005}, infrared light \citep[e.g.][]{Drimmel2000, DrimmelSpergel2001}, and dust maps \citep[e.g.][]{Rezaei2018}. Various studies aimed to quantify the mass density contrast between the arm and inter-arm regions of extragalactic spiral galaxies \citep{Rix1995, Meidt2012, Meidt2014} as well as the the Milky Way \citep{Siebert2012}, the pitch angle of the spiral arms \citep[e.g.][]{Vallee2015}, as well as their pattern speed \citep{Fernandez2001, Kranz2003, Martos2004, DiasLepine2005, Gerhard2011}. 

In all of these studies it is conceptually advisable, but practically difficult, to differentiate between the dynamical non-axisymmetric (spiral) perturbation that `drives' the dynamics, and the spiral-like morphology in gas, dust and young stars distribution that -- at least in part -- resulted from and merely traces this dynamical perturbation. 

Our position within the Milky Way close to the Galactic mid-plane makes it very challenging to directly map the large-scale morphology of spiral arm tracer, let alone any spiral {\it mass density} structure of the Galactic disk, due to line-of-sight effects, distance uncertainties, and interstellar extinction \citep[e.g.][]{Schlegel1998, Schlafly2014, Rezaei2018}. Additionally, various stellar surveys have different depths, and often only loosely or undetermined selection functions, making a comparison between different data sets difficult. As a result no consensus has yet been achieved in the literature on the number of spiral arms in our Galaxy, their location, or density contrast. 

While many of the previous studies were focused on tracers of star forming regions or young stars born within the spiral arms, all disk stars can cross the arms and contribute to the stellar overdensities. Therefore all stellar populations will contribute to a dynamical spiral perturbation, or disk surface mass density perturbation, in the Milky Way disk. Such a dynamical perturbation, or its corresponding perturbation in the Galaxy's gravitational potential, would imprint a non-axisymmetric signature on the kinematics of the disk stars \citep[e.g.][]{Monari2016}. Exploring evidence for such a signature, modelling it and thereby constraining any dynamical spiral in the Galactic disk is the focus of this paper.

Recently, the second data release (DR2) of the \gaia\ mission has enabled the largest data set to date with $6$D phase-space information (position and velocities) for millions of stars, ushering us in a new era of Galactic astronomy \citep{Gaia2016}. However, while the astrometric precision of the data set is unprecedented, uncertainties in the parallax estimates significantly dominate the error budget already beyond a few kpc distance from the Sun. 
Thus directly mapping overdensities in the stellar disk by star counts remains challenging for two main reasons: firstly, the complex and unknown selection function of the observed stars impedes a full census of all disk stars, and second, the imperfect distance information dilutes spatial density structure. 

We have recently addressed the latter problem by developing a data-driven model to determine precise parallax estimates for luminous red giant stars based on their multi-band photometry and spectroscopy, which is described in detail in \citet{HER2019}. These estimates enable us to construct global maps of the Milky Way out to large Galactocentric distances, i.e. $R\approx 25$~kpc. 
In this work, we circumvent the problem of the complex and unknown spatial selection function by studying the \textit{kinematics} of the observed stars, since any survey's selection function does not significantly depend on the objects' velocities. 

To this end, we present here the detection of a non-axisymmetric spiral pattern observed in the mean Galactocentric radial velocities $\langle v_R\rangle(R, \varphi)$ of luminous red giant stars within the Galactic mid-plane. 
Assuming a perturbation in the gravitational potential due to logarithmic spiral arms gives rise to this velocity pattern, we construct a simple steady-state toy model to interpret our observations. We constrain the density contrast between the arm and inter-arm regions and make predictions for stellar overdensities within the Milky Way's disk based on this model. 


\section{Data}\label{sec:data}

Our analysis is based on stars on the upper red giant branch (RGB), which are very luminous objects that can thus be observed out to large distances, covering a wide range of Galactocentric radii. 
In \citet{HER2019} we developed a data-driven method to infer precise parallaxes for these stars, assuming that red giant stars are dust-correctable, standardizable candles, which means that we can infer their distance modulus -- and thus their parallax -- from their spectroscopic and photometric features. Our method employs a purely linear function of spectral pixel intensities from \apogee\ DR14 \citep{Majewski2017}, as well as multi-band photometry from \gaia\ DR2 \citep{Gaia2016}, \zmass\ \citep{2MASS}, and \wise\ \citep{WISE}. To this end, we restrict ourselves to a limited region of the stellar parameter space, and analyze only stars with very low surface-gravity, i.e. $0\leq \log g \leq 2.2$, which selects stars that are more luminous than the red clump. 

Our model predicts spectrophotometric parallaxes for $\sim 45,000$ RGB stars with uncertainties better than $\sim 10\%$, which results in distance estimates that are more than twice as precise as \gaia's predictions at heliocentric distances of $\gtrsim 3$~kpc ($\gtrsim 1$~kpc) for stars with $G\sim 12$~mag ($G\sim14$~mag). At $\approx 15$~kpc distance from the Sun, our derived spectrophotometric distances have $\sim 6-8$ times smaller uncertainties than the distances from astrometric parallaxes reported by \gaia\ DR2 and thus this data set with complete $6$D phase-space information allows us to make global maps of the Milky Way \citep{Eilers2019, HER2019}. 

We transform all stars to the Galactocentric cylindrical coordinate frame making use of the barycentric radial velocities derived by \apogee\ DR14 and precise proper motions delivered from \gaia\ DR2. The Galactocentric azimuth angle $\varphi$ is measured from the centre--anticentre line with $\varphi$ increasing counter-clockwise, i.e. in the opposite direction of Galactic rotation. We assume a distance from the Sun to the Galactic center of $R_{\odot}=8.122\pm0.031$~kpc \citep{gravity2018}, a height of the Sun above the Galactic plane of $z_{\odot}\approx 0.025$~kpc \citep{Juric2008}, and the Galactocentric velocity components of the Sun $v_{\odot,\,x}\approx-11.1\,\rm km\,s^{-1}$, $v_{\odot,\,y}\approx245.8\,\rm km\,s^{-1}$, and $v_{\odot,\,z}\approx7.8\,\rm km\,s^{-1}$, which have been derived from the proper motions of $\rm Sgr~A^{*}$ \citep{ReidBrunthaler2004}. 
The derived velocity components for each star in Galactic cylindrical coordinates $v_R$, $v_{\varphi}$, and $v_z$ have median uncertainties of $\sigma_{v_R} \approx 1.9\,\rm km\,s^{-1}$, $\sigma_{v_{\varphi}} \approx 2.5\,\rm km\,s^{-1}$, and $\sigma_{v_z} \approx 2.1\,\rm km\,s^{-1}$, which we estimated via Monte Carlo sampling.

\section{Observed Kinematic Spiral Signature}\label{sec:signature}

We restrict our analysis to all stars that lie within $|z|\leq0.5$~kpc distance from the Galactic mid-plane, or within $6^{\circ}$, i.e. $|z|/R \leq \tan(6^{\circ})$, to account for the flaring of the disk. We cut down our sample based on vertical velocities of $|v_z| < 100\,\rm km\,s^{-1}$ in order to eliminate halo stars. Furthermore, we select stars with low $\alpha$-element abundances, i.e. $[\alpha/\rm Fe] < 0.12$, to avoid large asymmetric drift corrections \citep[e.g.][]{Golubov2013}. This results in $32,271$ RGB stars within the Milky Way's disk, covering distances ranging from the Galactic center out to $R\approx 25$~kpc. 

\begin{figure}[!t]
\includegraphics[width=\textwidth]{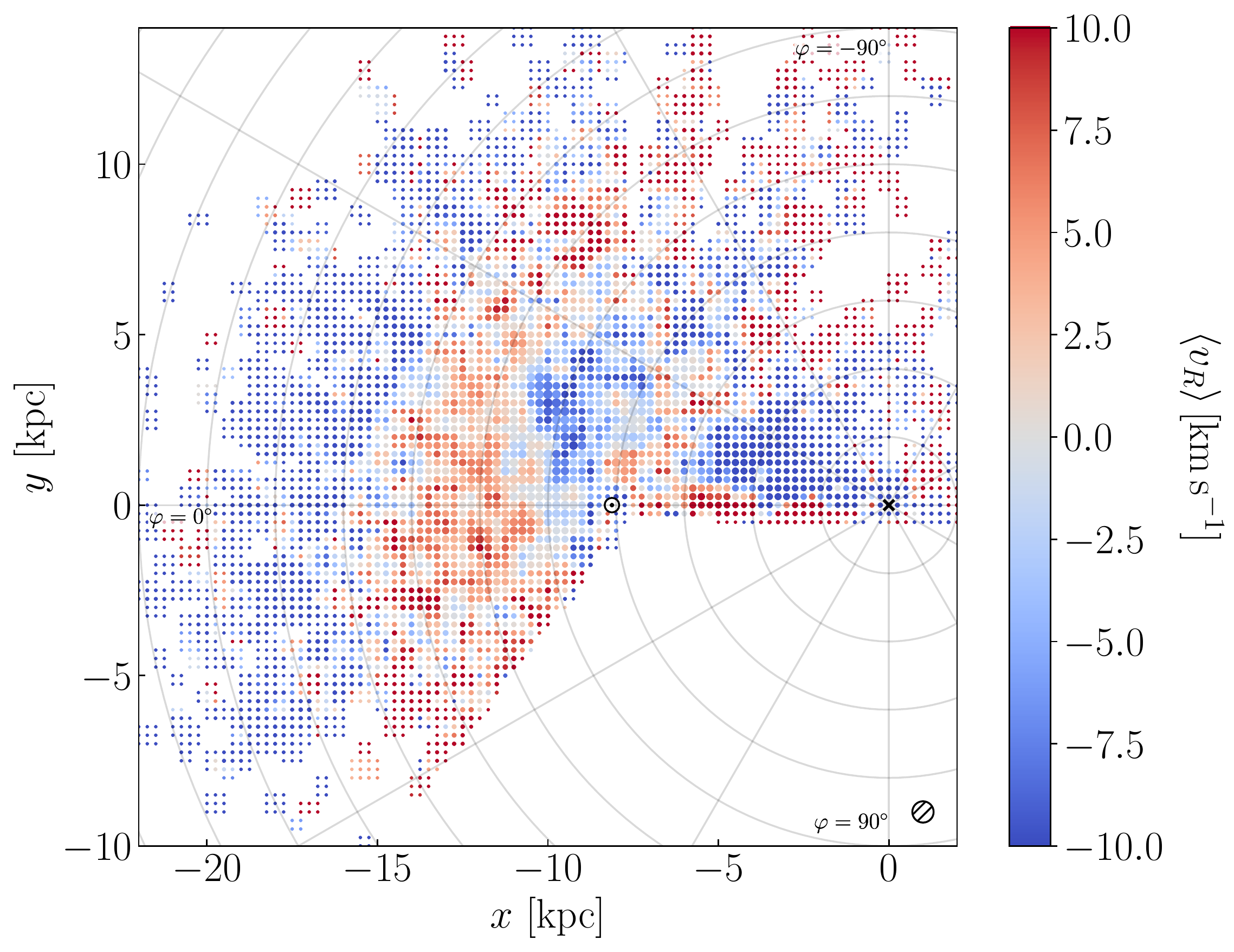}
\caption{\textbf{Kinematic spiral signature within the Milky Way's disk.} The RGB stars within the Milky Way's mid-plane are shown averaged in spatial bins with $0.25$~kpc on a side, colored by their mean Galactocentric radial velocity. Note that in practice we average the radial velocity of all stars within $2.5$ times the size of each bin. The ``effective'' resolution of the map is indicated by the hashed circle in the lower right corner. The locations of the Galactic center and the Sun are indicated by the black cross and the $\odot$ symbol, respectively. The light grey curves mark concentric circles at $\Delta R=2$~kpc, as well as azimuthal angles of $\Delta \varphi=30^{\circ}$. \label{fig:data}}
\end{figure}

In Fig.~\ref{fig:data} we show these stars in bins of $0.25$~kpc on a side in $x-$ and $y-$direction. The data points are colored by the mean Galactocentric radial velocity $\langle v_R\rangle$ of all stars within each bin and the size of the points reflects the number of stars. Groups of stars moving on average towards the Galactic center are colored in blue, whereas outwards moving stars are colored in red. The resulting pattern in the mean radial velocities of the disk stars shows a spiral signature that we would not expect to see in an unperturbed axisymmetric gravitational potential. 

Note that we average the radial velocity of all stars within $2.5$ times the size of each bin, i.e. our map has an ``effective'' resolution of $\Delta x = \Delta y = 0.625$~kpc, which we chose in order to reveal the spiral signature more clearly. This introduces correlations between the different bins, and thus the individual bins cannot be treated statistically independently anymore. 

\begin{figure}[!t]
\includegraphics[width=\textwidth]{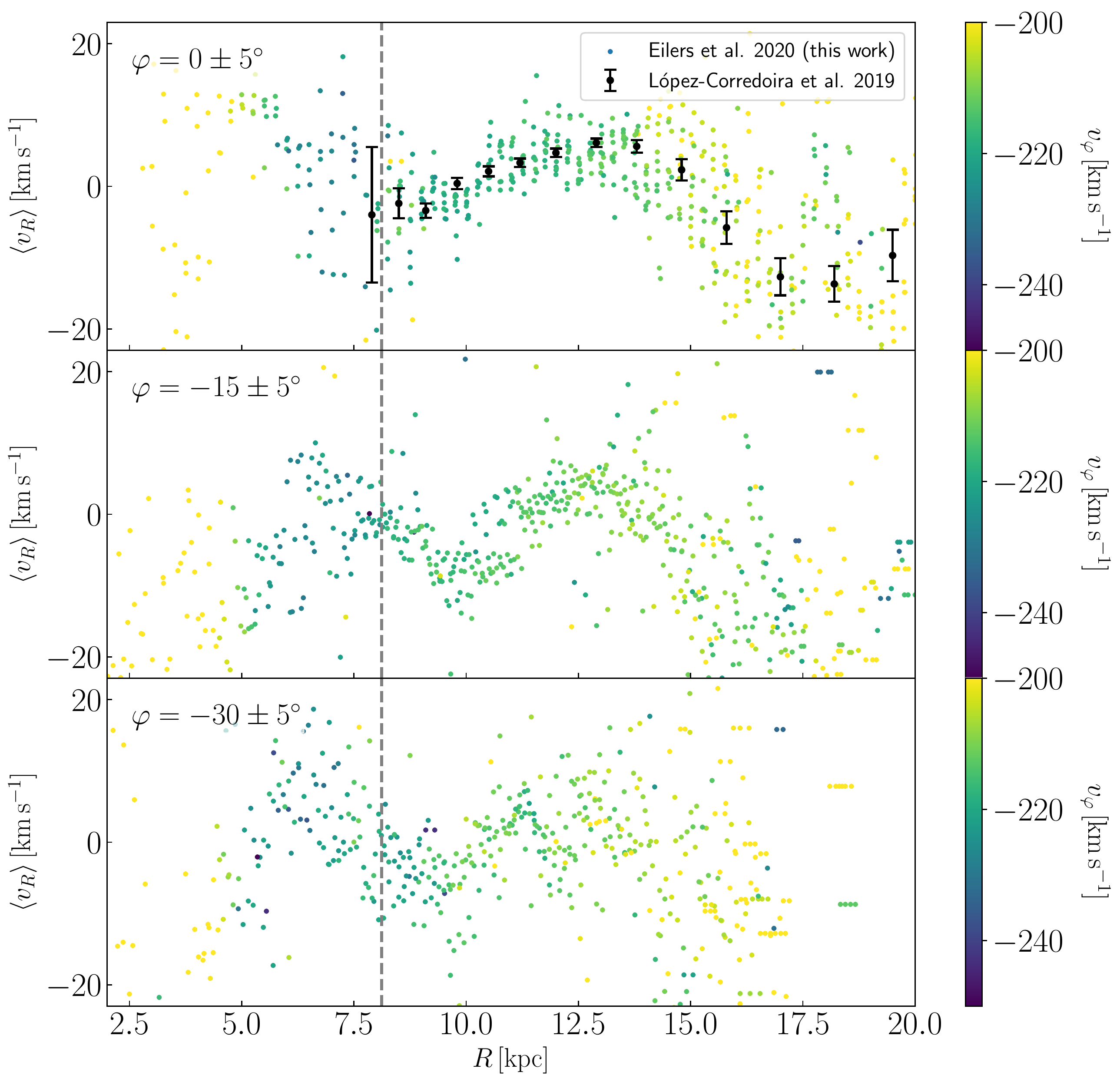}
\caption{\textbf{Radial profile of the mean radial velocity at different azimuth angles.} Data points show the averaged values within the same bins as in Fig.~\ref{fig:data}, colored by the mean rotational velocity $v_{\varphi}$. The different panels show thin ($10^{\circ}$) wedges in azimuth. The grey dashed lines mark the solar radius $R_{\odot}$. The top panel also shows measurements by \citet{LopezCorredoira2019}. \label{fig:data_vr}}
\end{figure}

The same velocity signature can also be seen in Fig.~\ref{fig:data_vr}, where we show the mean radial velocity as a function of Galactocentric radius $R$ for wedges of $10^{\circ}$ at different azimuth angles $\varphi$. The mean radial velocity oscillates around $v_R\sim0\,\rm km\,s^{-1}$. 

The inner part of our Galaxy is dominated by the bulge with a central bar. \citet{Bovy2019} showed that the Milky Way's bar introduces a quadrupole moment in the radial velocities of stars in the central region of our Galaxy. Our data reveal a similar feature within $R\lesssim 5\rm\, kpc$ as seen in Fig.~\ref{fig:data}, although the angle of the quadrupole moment in our data is misaligned with what is currently believed to be the major-axis angle of the Galactic bar by $\sim 20^{\circ}$ \citep{Wegg2015}. This could result from systematic uncertainties in the distance estimates towards the Galactic center, where crowding and high dust extinction might introduce biases on the spectrophotometric parallax estimates. Alternatively, the mismatch between the quadrupole moment and the Milky Way's bar could be real, possibly induced by a recent interaction with the Sagittarius dwarf galaxy, as recently suggested by \citet{Carrillo2019}. Further investigation will be necessary to securely determine the significance and potential reason for this offset. However, since our study focuses on the observed velocity pattern outside the Galactic bulge, we will postpone this question to future work. 

\section{Steady-State Toy Model}\label{sec:model}

We now present the arguably simplest, but also simplistic model of the observed velocity field. To this end, we model the loop orbits of stars in a gravitational potential as a superposition of a guiding center and small oscillations around this guiding center, following the derivation of \citet[][see \S~$3.3.3$]{BinneyTremaine}, who compute the effects of a perturbation in the gravitational potential due to a weak bar. We neglect any vertical velocity components and only model the kinematics within the Galactic mid-plane. 

We begin with the equations of motions in polar coordinates obtained from the Lagrangian function, i.e. 
\begin{align}
    \ddot{R} &= R(\dot{\phi} + \Omega_{\mathrm{p}})^2 - \frac{\partial\Phi}{\partial R}; \label{eq:motion1}\\
    \frac{\rm d}{{\rm d}t}\left[R^2(\dot{\phi} + \Omega_{\mathrm{p}})\right] &= - \frac{\partial\Phi}{\partial \phi}, \label{eq:motion2}
\end{align}
where $\Omega_{\mathrm{p}}$ describes the rotation frequency of the pattern speed of what will be a rotating perturbation, and $\phi$ describes the azimuthal angle in the co--rotating frame, i.e $\phi = \varphi - \Omega_{\mathrm p}t$. 
We then introduce a non-axisymmetric rotating perturbation in the gravitational potential, which in turn perturbs the radius and azimuthal angle of the stars: 
\begin{align}
    \Phi(R, \phi) &= \Phi_0(R) + \Phi_1(R, \phi);\\
    R(t) &= R_0 + R_1(t);\\
    \phi(t) &= \phi_0(t) + \phi_1(t). 
\end{align}
The index $0$ denotes the unperturbed quantity, where as an index $1$ indicates the (small) perturbation. 

This perturbation in the gravitational potential leads to the following first-order terms in the equations of motion: 
\begin{align}
    \ddot{R}_1 + \left(\frac{{\rm d}^2\Phi_0}{{\rm d}R^2}-\Omega^2\right)_{R_0} R_1-2R_0\Omega_0\dot{\phi}_1 &= -\left(\frac{\partial\Phi_1}{\partial R}\right)_{R_0};\label{eq:ddotR1}\\
    \ddot{\phi}_1 + 2\Omega_0\frac{\dot{R}_1}{R_0} &= -\frac{1}{R_0^2}\left(\frac{\partial\Phi_1}{\partial\phi}\right)_{R_0}\label{eq:ddotphi1}, 
\end{align}
where we introduced the circular frequency $\Omega(R) = \sqrt{\frac{1}{R}\frac{{\rm d}\Phi_0}{{\rm d}R}}$, with $\Omega_0 = \Omega(R_0)$. 

We now choose a specific form of the perturbing potential, assuming it arises due to a logarithmic spiral \citep{Kalnajs1971, BinneyTremaine}, i.e. 
\begin{align}
 \Phi_1(R, \phi) = A(R) \, \exp\left[i\left(m \phi + m\,\frac{\ln(R/h_{R,1})}{\tan p}\right)\right], \label{eq:spiral_pot}
\end{align}
where $m$ and $p$ determine the number and pitch angle of the spiral arms, respectively, $h_{R,1}$ is the scale length of the potential perturbation, and $A(R)$ is an amplitude.  

We then insert this perturbing potential into Eqns.~\ref{eq:ddotR1} and \ref{eq:ddotphi1}, and integrate the latter once to obtain $\dot{\phi}_1$, which will be replaced in Eqn.~\ref{eq:ddotR1}. We assume $\phi_1\ll 1$, which is a valid assumption in the absence of resonances (see \S~\ref{sec:more_model}). This implies $\phi\approx \phi_0$, and $\varphi\approx\varphi_0=\phi_0+\Omega_{\mathrm p}t$. 

Taking the real part of the resulting Eqn.~\ref{eq:ddotR1}, we obtain 
\begin{align}
    \ddot{R}_1 +\kappa_0^2 R_1 = -\left[\frac{2\Omega_0A}{R_0(\Omega_0-\Omega_{\mathrm{p}})}+\left(\frac{{\rm d} A}{{\rm d} R}\right)_{R_0}\right]\,\cos\left[m\left(\varphi_0-\Omega_{\mathrm{p}}t\right)+m\,\frac{\ln(R_0/h_{R,1})}{\tan p}\right] \label{eq:oscillator}, 
\end{align}
with the epicycle frequency
\begin{align}
    \kappa_0^2 = \left(R\frac{{\rm d}\Omega^2}{{\rm d}R} +4\Omega^2\right)_{R_0}. 
\end{align}
Eqn.~\ref{eq:oscillator} describes a driven harmonic oscillator which has a solution 
\begin{align}
    R_1(t) &= C_0 \cos\left[\kappa_0t+\alpha\right] \nonumber \\
    &- \left[\frac{2\Omega_0A}{R_0(\Omega_0-\Omega_{\mathrm{p}})}+\left(\frac{{\rm d}A}{{\rm d}R}\right)_{R_0}\right]\,\frac{1}{\Delta}\, \cos\left[m(\varphi_0 - \Omega_{\mathrm{p}}t) + m\,\frac{\ln(R_0/h_{R,1})}{\tan p}\right]\label{eq:solution}, 
\end{align}
where $C_0$ and $\alpha$ are constants, and $\Delta = \kappa_0^2-m^2(\Omega_0-\Omega_{\mathrm{p}})^2$. 
The radial velocity can now be derived via $v_R = \dot{R}(t) = \dot{R}_1(t)$, i.e.
\begin{align}
    v_R &= -C_0 \kappa_0\sin\left[\kappa_0t+\alpha\right] \nonumber \\
    &+ \left[\frac{2\Omega_0A}{R_0(\Omega_0-\Omega_{\mathrm{p}})}+\left(\frac{{\rm d}A}{{\rm d}R}\right)_{R_0}\right]\,\frac{m(\Omega_0-\Omega_{\mathrm{p}})}{\Delta}\, \sin \left[m(\varphi_0 - \Omega_{\mathrm{p}}t) + m\,\frac{\ln(R_0/h_{R,1})}{\tan p}\right].\label{eq:vR}
\end{align}

Assuming that the level of potential perturbation is roughly proportional to the level of density perturbation, the surface density perturbation $\Sigma_1(R, \varphi)$ can be derived by means of the Poisson equation for a disk within the Galactic mid-plane, i.e. 
\begin{equation}
    \nabla^2\tilde{\Phi}_1(R, \varphi, z) = 4\pi\,G\,\Sigma_1(R, \varphi) \, \delta(z), \label{eq:poisson}
\end{equation}
where $\tilde{\Phi}_1(R, \varphi, z)$ is the $3$D potential perturbation, and $\delta(z)$ is the Dirac delta function. We approximate the $3$D gravitational potential perturbation by 
\begin{equation}
\tilde{\Phi}_1 (R, \varphi, z) = \Phi_1(R, \varphi)\, \exp(-|z|/h_z), 
\end{equation}
where we introduced the disk's scale height $h_z$ \citep[see][\S~2.6]{BinneyTremaine}. 
Integrating Eqn.~\ref{eq:poisson} along the $z$-direction from $z=-\infty$ to $z=\infty$, 
we obtain: 
\begin{equation}
    \frac{\partial^2\Phi_1(R, \varphi)}{\partial R^2} + \frac{1}{R}\frac{\partial \Phi_1(R, \varphi)}{\partial R} + \frac{1}{R^2}\frac{\partial^2 \Phi_1(R, \varphi)}{\partial\varphi^2} = \frac{2\pi G \Sigma_1(R, \varphi)}{h_z}. \label{eq:poisson2D}
\end{equation}
We only consider the dominant term of the left side of Eqn.~\ref{eq:poisson2D}, which is proportional to $\tan^{-2} p$, assuming the spiral is tightly wound, i.e. the pitch angle is ``small''. This Wentzel–Kramers–Brillouin (WKB) approximation is sensible for pitch angles $p<15^\circ$ \citep[see][\S~6.2.2]{BinneyTremaine}

Introducing the maximum surface density within the spiral arms, i.e. 
\begin{equation}
\Sigma_{\rm max}(R) = \Sigma_{\rm max}(R_{\odot})\,\exp\left[-\frac{R-R_{\odot}}{h_{R, 1}}\right],  
\end{equation}
the amplitude $A(R)$ of the potential perturbation can then be expressed in terms of 
\begin{equation}
    A(R) = -\frac{2\pi\,G\,\Sigma_{\rm max}(R)\,R^2\,\tan^2 p}{h_z\,m ^2}. 
\end{equation}
Thus we obtain a surface density perturbation of
\begin{equation}
    \Sigma_1(R, \varphi) = \Sigma_{\rm max}(R) \, \cos\left[m\left(\varphi -\Omega_{\mathrm{p}} t\right) + m\,\frac{\ln(R/h_{R,1})}{\tan p }\right],  \label{eq:Sigma1}
\end{equation}
assuming that the amplitude of the surface density perturbation is proportional to the level of potential perturbation. 

Note that while the surface density perturbation declines on a scale length $h_{R, 1}$, the unperturbed surface density $\Sigma_0(R)=\Sigma_0(R_{\odot})\,\exp\left[-\frac{R-R_{\odot}}{h_{R, 0}}\right]$ declines on the Galactic disk scale length $h_{R, 0}$. 

\subsection{Modeling the Observed Kinematic Spiral Signature}\label{sec:more_model}

\begin{figure}[!t]
\centering
\includegraphics[width=.49\textwidth]{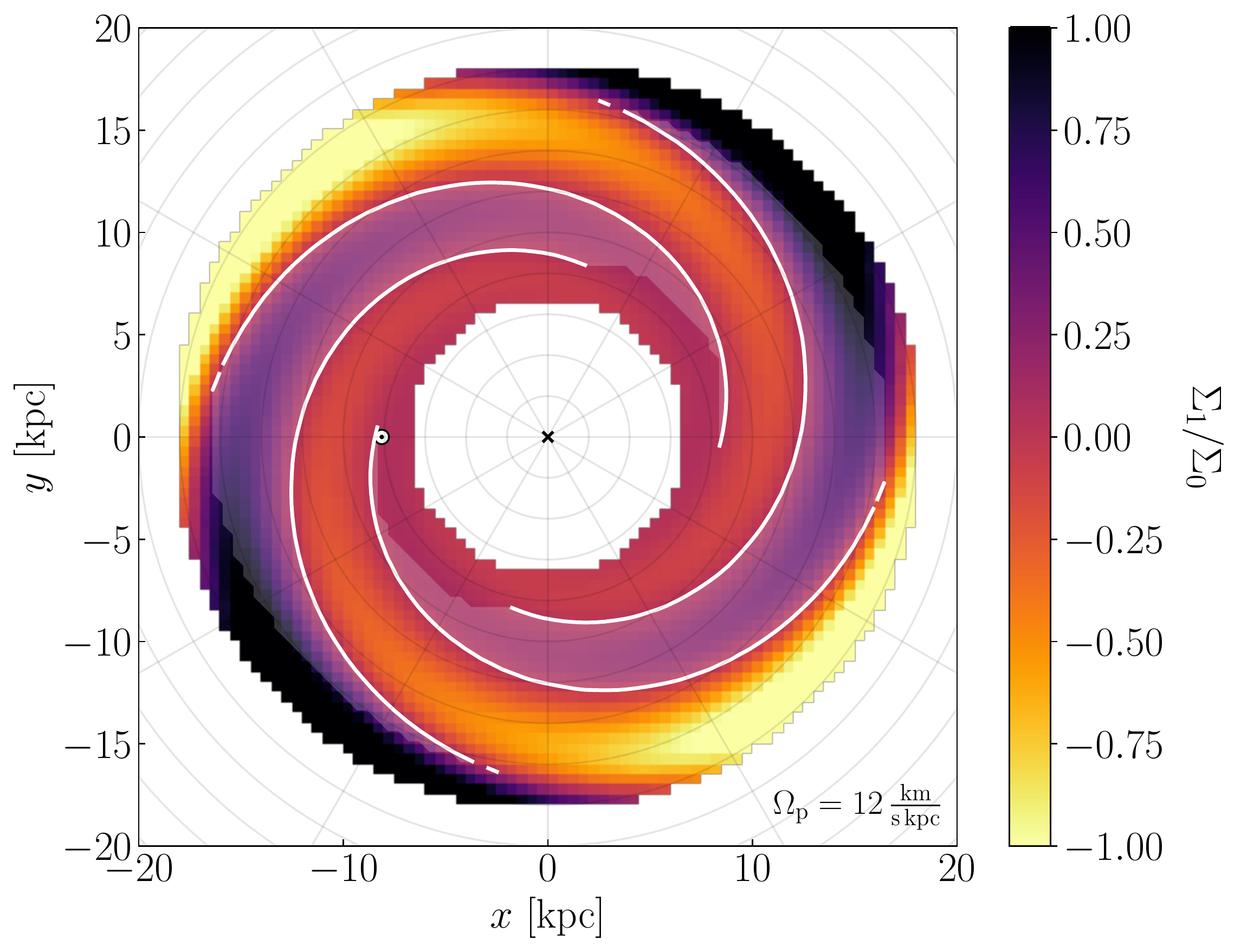}
\includegraphics[width=.49\textwidth]{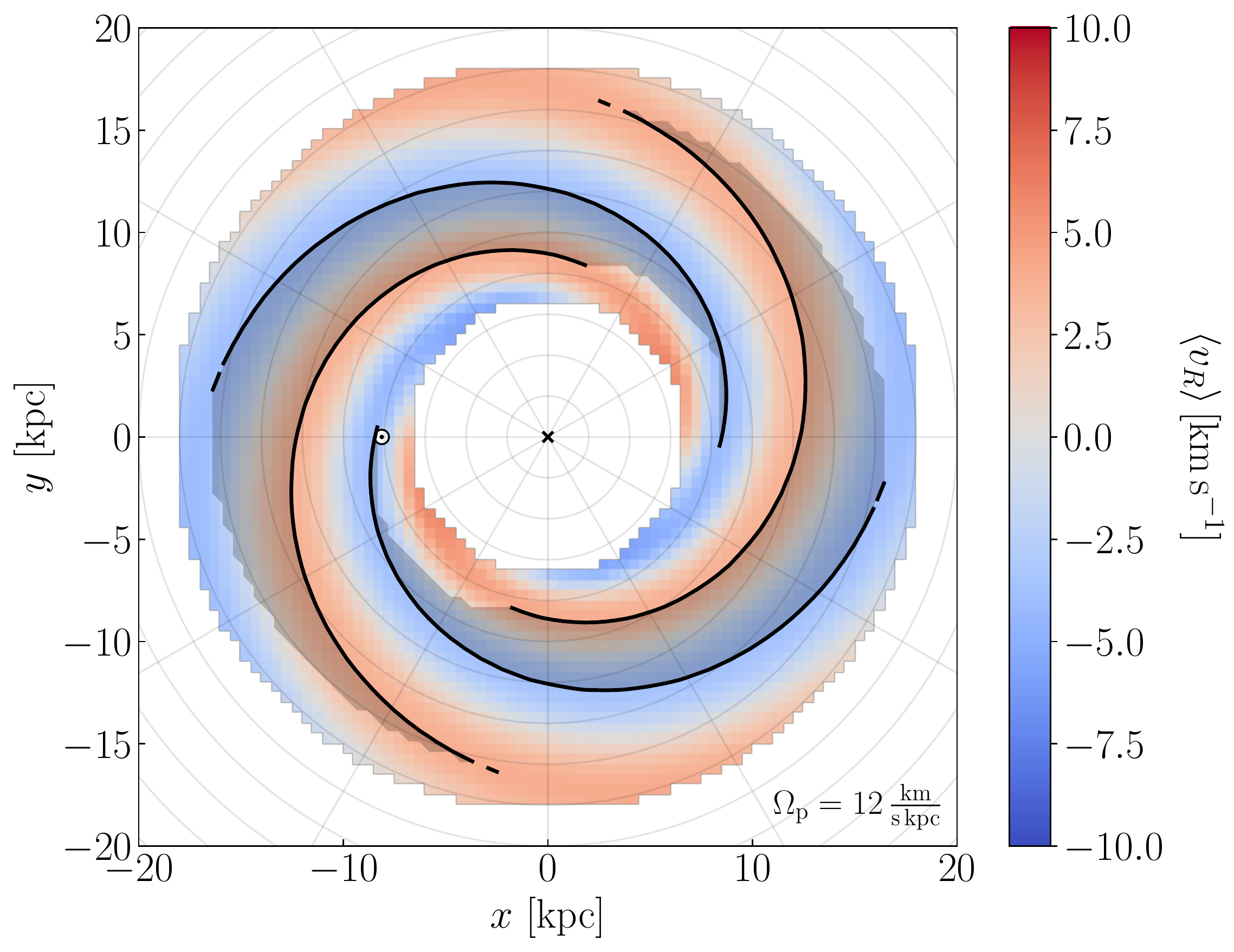}\\
\includegraphics[width=.49\textwidth]{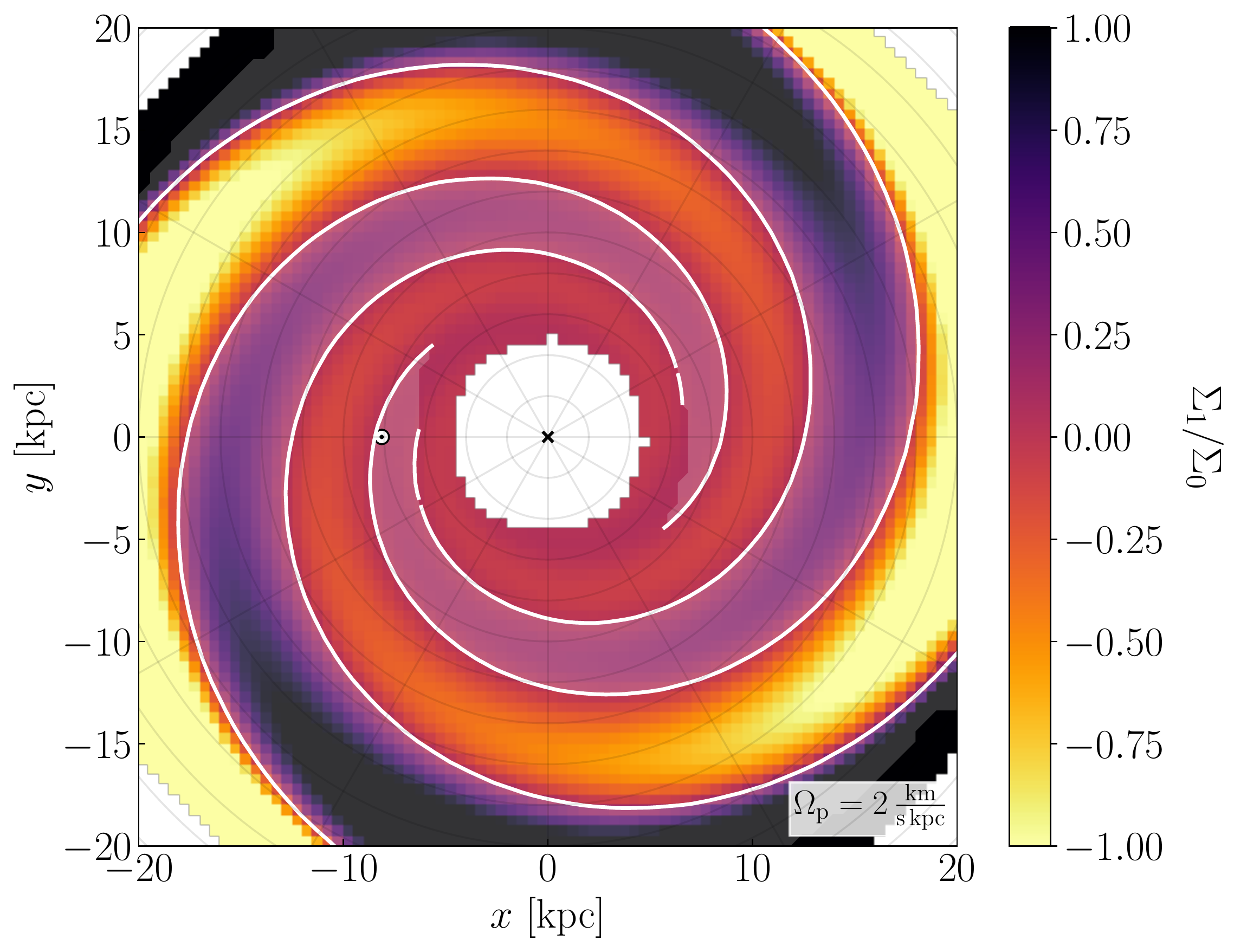} 
\includegraphics[width=.49\textwidth]{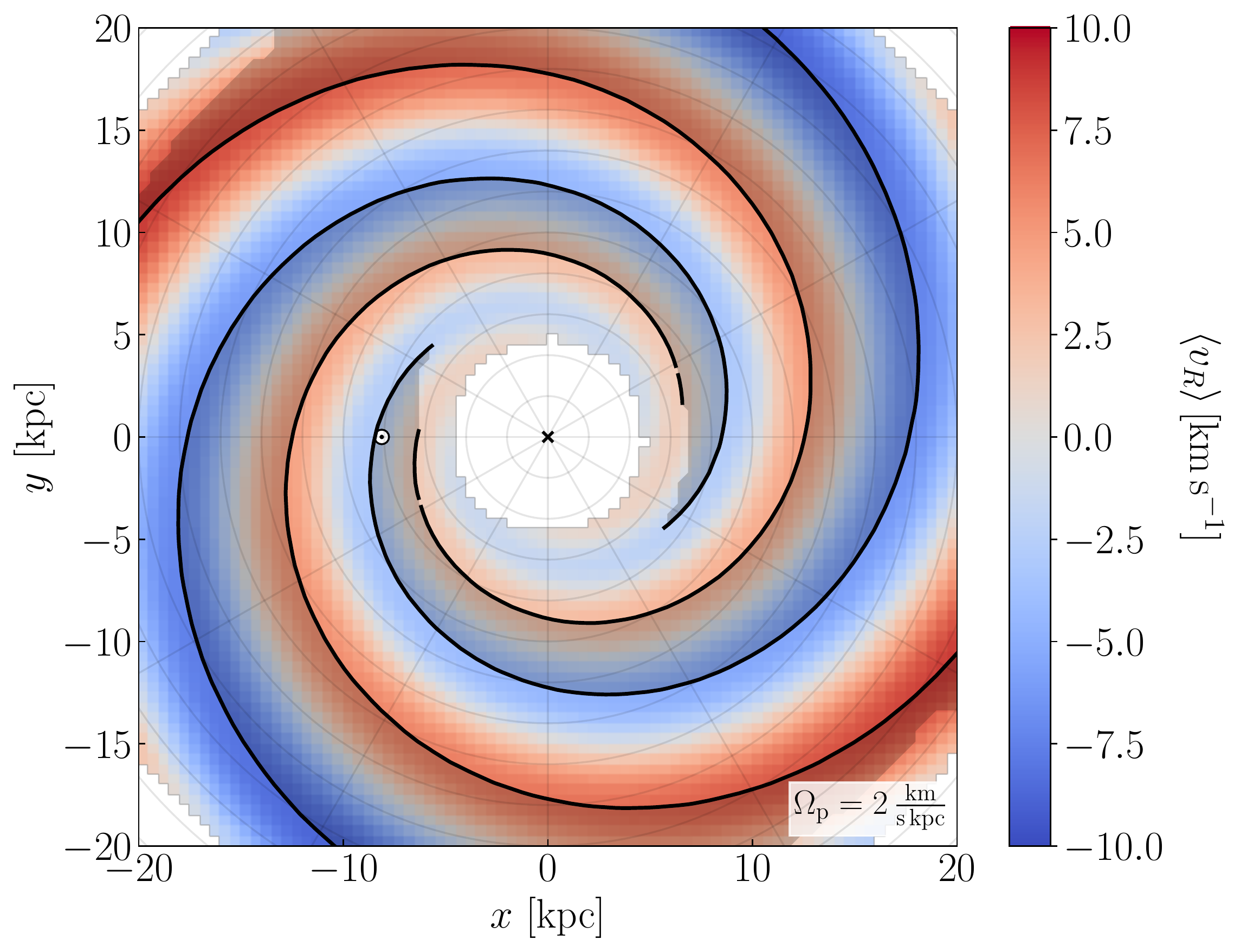}
\caption{\textbf{Toy model of a non-axisymmetric perturbation in a gravitational potential due to logarithmic spiral arms.} \textit{Left panels:} Modeled surface density perturbation contrast $\Sigma_1/\Sigma_0$ in the disk showing a two armed spiral pattern. The white contours trace the overdensities. \textit{Right panels:} Radial velocity signature introduced by the potential perturbation with the same surface density contours (here in black). The top panels show our best model with a chosen pattern speed of $\Omega_{\mathrm{p}}=12\,\rm km\,s^{-1}\,kpc^{-1}$ with stars at $7<R<18$~kpc used to constrain the model, whereas the bottom panels show our best model with a smaller pattern speed of $\Omega_{\mathrm{p}}=2\,\rm km\,s^{-1}\,kpc^{-1}$ fitted to all stars within $5<R<25$~kpc. \label{fig:model}}
\end{figure}

In order to provide a qualitative model of the observed spiral signature from Fig.~\ref{fig:data}, we choose $C_0=0$ (i.e. closed loop orbits), and a perturbation arising from a two-armed logarithmic spiral, i.e. $m=2$. 
Furthermore, we assume a flat circular velocity curve with $v_{\rm c}(R)=229~\rm km\,s^{-1}$ at all radii \citep{Eilers2019}. 

We chose two different pattern speeds: first, we chose a very small pattern speed of $\Omega_{\mathrm{p}}=2\,\rm km\,s^{-1}\,kpc^{-1}$, which avoids all resonances, i.e. the resonance at the co-rotation radius as well as the inner (and outer) Lindblad resonances. For this pattern speed, all of our data lie within the inner Lindblad resonance. Thus we use $24,923$ stars in our data set at Galactocentric radii of $5\leq R \leq 25$~kpc to constrain this model, only avoiding the inner part of the disk dominated by the Galactic bulge, which is not included in the model. However, it has been shown that most spiral structure lies in between the inner Lindblad resonance and the co-rotation radius \citep[e.g.][]{SellwoodBinney2002}, and hardly any spirals exist inside the inner Lindblad resonance. Consequently, the choice of $\Omega_{\mathrm{p}}=2\,\rm km\,s^{-1}\,kpc^{-1}$, serving us by avoiding all resonances in the radial range of the data, has the drawback that it places the adopted spiral in a radius range, where self-consistent dynamics indicate spirals should not live. Therefore, we also choose a far larger pattern speed of $\Omega_{\mathrm{p}}=12\,\rm km\,s^{-1}\,kpc^{-1}$. This pattern speed is more realistic, placing the spiral between the inner Lindblad resonance and co-rotation where dynamics says spirals live. However, this forces us to restrict our data to stars within $7\leq R\leq 18\,\rm kpc$, in order to avoid these resonances. This data set is only slightly smaller, i.e. it contains $22,924$ stars, since most of the observed disk stars are located with this region. 
Furthermore, we chose a disk scale height of $h_z=1$~kpc. 

Our model now has four remaining free parameters, i.e. the rotation angle of the spiral pattern which is given by the time $t$, the pitch angle of the spiral arms $p$, the maximum surface density of the perturbation at the solar radius $\Sigma_{\rm max}(R_{\odot})$, and the scale length of the perturbation $h_{R,1}$. We optimize our model to obtain the best estimates for these free parameters by means of a least square minimization making use of the Markov Chain Monte Carlo algorithm \texttt{emcee} \citep{emcee} with flat priors of $h_{R,1}\in[1, 50]$~kpc, $\Sigma_{\rm max}(R_{\odot})\in[0, 50]\,\rm M_{\odot}\,pc^{-2}$, $p\in[0.1, 0.3]$ and $t\in[6, 8]$~Gyr. 

Our best model estimates for a pattern speed of $\Omega_{\mathrm{p}}=12\,\rm km\,s^{-1}\,kpc^{-1}$ ($\Omega_{\mathrm{p}}=2\,\rm km\,s^{-1}\,kpc^{-1}$) indicate an amplitude of $\Sigma_{\rm max}(R_{\odot})= 5.48\pm0.01\,\rm M_{\odot}\,pc^{-2}$ ($\Sigma_{\rm max}(R_{\odot})= 6.00\pm0.03\,\rm M_{\odot}\,pc^{-2}$), and a pitch angle $p=0.2101\pm0.0002$ ($p=0.2210\pm0.0002$), when evolving the system until a time $t=6.95\pm0.01$~Gyr ($t=7.47\pm0.01$~Gyr). However, given the rotational symmetry the model gives rise to the same pattern when rotated by $\pi$. For the scale length of the perturbation we obtain $h_{R, 1}\gtrsim50$~kpc ($h_{R, 1}= 11.36\pm 0.06$~kpc), which indicates that the amplitude of the perturbation stays approximately constant within the extent of the disk that is covered by our data. Note that the statistical uncertainties on the model parameters are very small, but the dominant uncertainty on our best fit parameters comes from the systematic errors associated with the choice of our toy model. 

Our resulting best models are shown in Fig.~\ref{fig:model}. The left panels shows the surface density perturbation $\Sigma_1$ divided by the unperturbed surface density $\Sigma_0$ of the disk stars assuming a scale length of $h_{R,0}=3$~kpc and a surface density at the solar radius of $\Sigma_0(R_{\odot})=68\,\rm M_{\odot}\,pc^{-2}$, which was determined by \citet{BovyRix2013}. The overdense regions of the two spiral arms are marked with white contours. 

\begin{figure}[!t]
\includegraphics[width=0.49\textwidth]{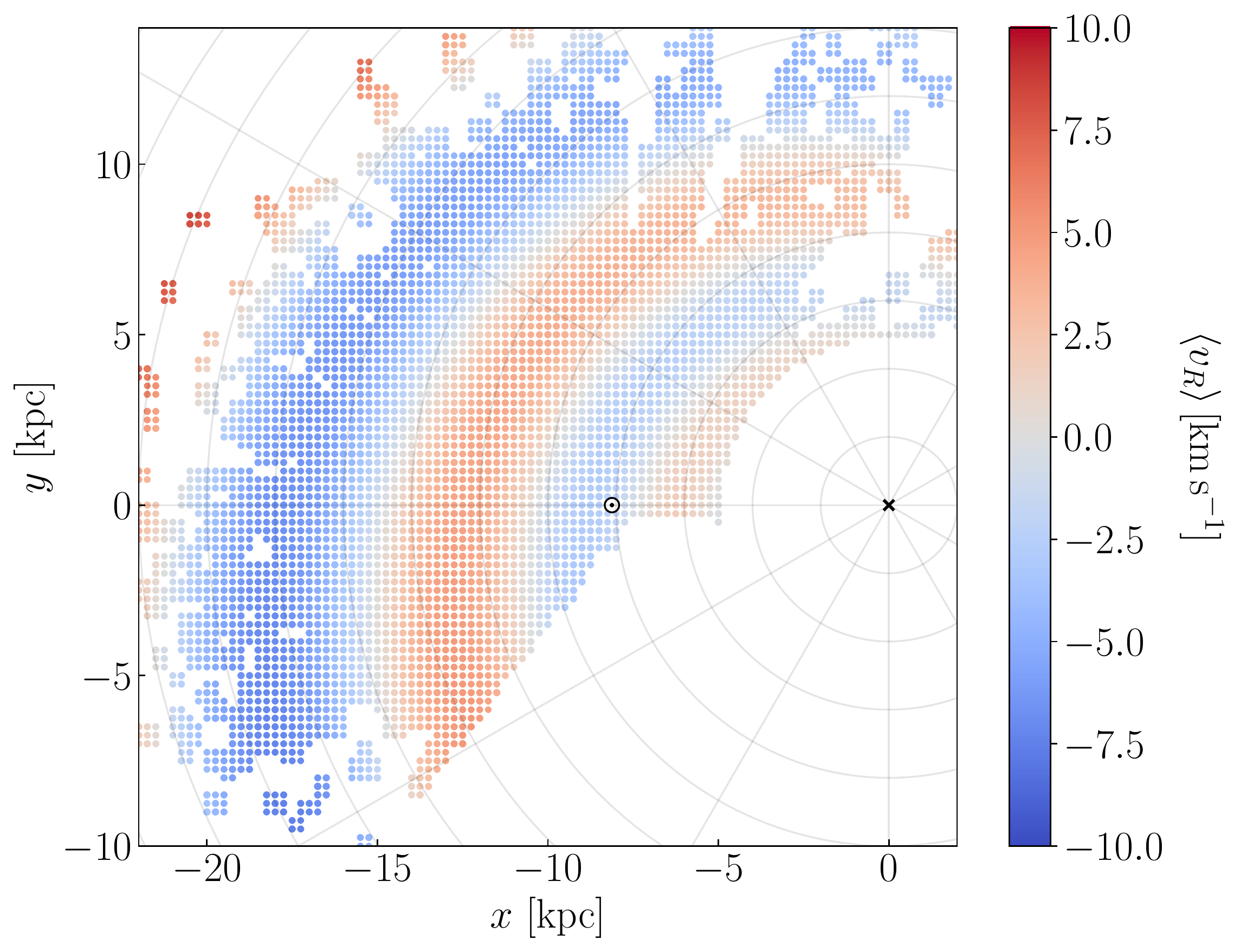}
\includegraphics[width=0.49\textwidth]{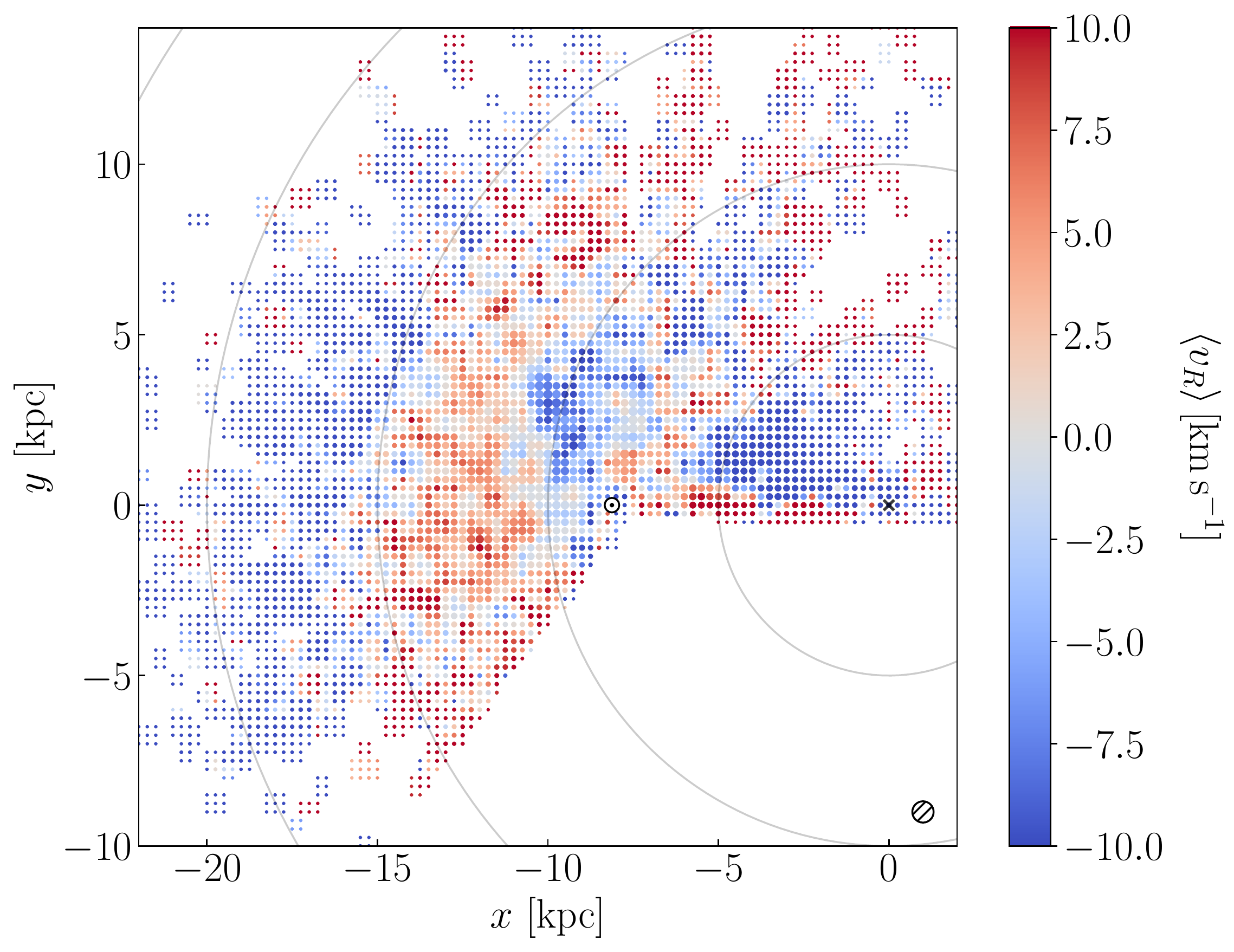}
\caption{Comparison of the radial velocity maps from our toy model with $\Omega_{\mathrm{p}}=2\,\rm km\,s^{-1}\,kpc^{-1}$ (\textit{left}) to the observations (\textit{right}) on the same spatial coverage. The inner part of the disk, i.e. $R\lesssim5$~kpc, is dominated by the Galactic bulge, and has thus been excluded from the model. \label{fig:model_prediction}}
\end{figure}

The right panels presents the mean radial velocity of the stars in our model, and the same contours tracing the overdensities on top in black. The radial velocity of the stars shows the same spiral pattern, although phase shifted with respect to the surface density perturbation. This phase shift is expected considering that the gravitational pull of an overdensity causes stars located at larger radii than the overdensity to move inwards, i.e. $\langle v_R\rangle < 0\,\rm km\,s^{-1}$, whereas stars at smaller radii will be accelerated outwards, i.e. $\langle v_R\rangle > 0\,\rm km\,s^{-1}$. 
In Fig.~\ref{fig:model_prediction} we show the best model estimate (for $\Omega_{\mathrm{p}}=2\,\rm km\,s^{-1}\,kpc^{-1}$, although the best model with $\Omega_{\mathrm{p}}=12\,\rm km\,s^{-1}\,kpc^{-1}$ looks qualitatively similar within the restricted radial range) on the same spatial coverage as the observations, revealing a spiral signature similar to our observations. 

Note that our model also makes predictions for the mean azimuthal velocity $\langle v_{\varphi}\rangle$. However, we have chosen not to conduct a comparison between the model and the observations for $\langle v_{\varphi}\rangle$, since the data looks very noisy once the mean circular velocity is subtracted. This is not surprising, since we have to subtract two large velocities off each other, making small velocity differences difficult to measure precisely. 

\begin{figure}[!t]
\includegraphics[width=\textwidth]{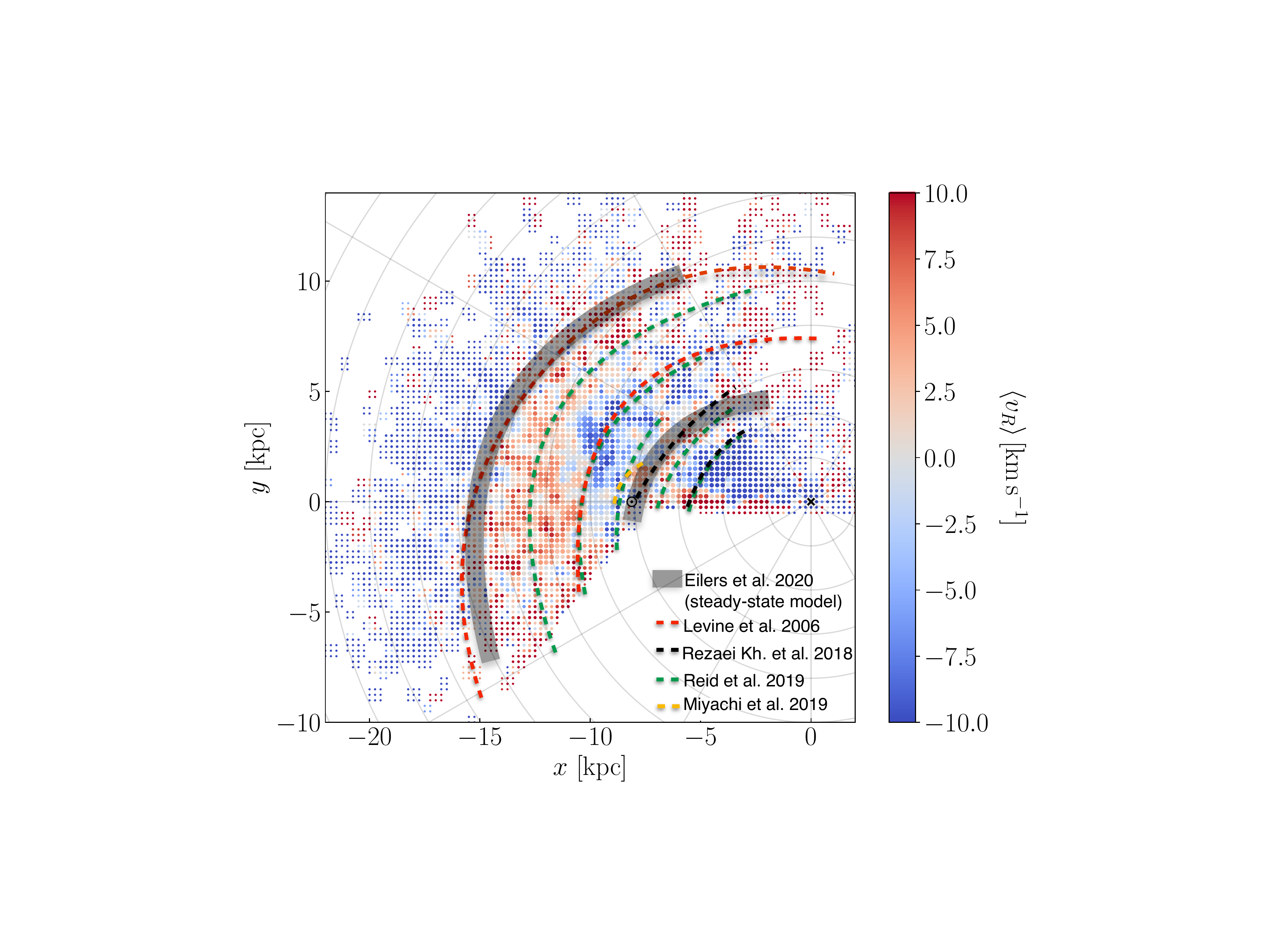}
\caption{\textbf{Comparison of the radial velocity map to predictions for the locations of spiral arms.} The black shaded regions indicate the locations of the overdensities predicted by our steady-state toy model, which trace the red-blue gradient in the velocity pattern. The red, black, green, and yellow colored dashed lines show predictions for the spiral arm locations by \citet{Levine2006}, \citet{Rezaei2018}, \citet{Reid2019}, and \citet{Miyachi2019}, respectively. \label{fig:comparison}}
\end{figure}

\section{Results}

Based on our steady-state toy model we can estimate the density contrast within the spiral arms and the inter-arm regions as well as make predictions for the locations of the spiral arms in the Milky Way. 

By combining the estimate of the maximum surface density perturbation with measurements of the local surface density from other studies we can constrain the density contrast between the arm and inter-arm regions of our Galaxy. Because we adopted $h_{R,0}=3$~kpc and obtain a much larger estimate of $h_{R,1}$, the density contrast between arm and inter-arm regions changes with radius. Assuming a surface density of $\Sigma_0(R_{\odot})=68\,\rm M_{\odot}\,pc^{-2}$ \citep{BovyRix2013} the density contrast at the solar radius is approximately $10\%$, with an increase towards larger radii. 


However, we have not explored how the estimated density contrast would change for a different choice for the functional form of the perturbation. If we had modeled the dynamical arms to be ``sharper'', e.g. by a phase-aligned superposition of a logarithmic two-- and four--arm spiral, the maximal surface density contrast (for a given potential perturbation) would be higher. 

Our steady-state toy model provides a prediction for the locations of the overdensities in the Milky Way, namely at the observed red--blue gradients in the radial velocity map, i.e. the transition from positive to negative velocities with increasing radius. At the location of these transitions, stars at smaller radii have on average larger radial velocities, whereas stars at larger radii have slightly smaller radial velocities. This causes all stars to approach the location of the red--blue gradients and thus resulting in an overdensity there, which we illustrate in Fig.~\ref{fig:comparison}. The predicted overdensities are approximately co-spatial with the location of the Local (Orion) Arm around $R\approx 8$~kpc and the Outer Arm in the outer part of the disk around $R\approx 15$~kpc \citep[e.g.][]{Levine2006, Camargo2015}. However, due to the likely transient nature of spiral arms the detailed relationship between the locations of overdensities and the velocity map will depend on whether the pattern is growing or decaying with time. 

We will now compare our results to several other studies that have analyzed the stellar kinematics and overdensities within the Milky Way's disk. 

\subsection{Comparison to Previous Studies}\label{sec:other_work}

Recently, \citet{LopezCorredoira2019} measured the radial profile of the Galactocentric radial velocity component of stars within the disk towards the Galactic anticenter at radii of $8\,{\rm kpc}\leq R\leq 28\,{\rm kpc}$ and $|z|<5\,\rm kpc$, likewise making use of \apogee\ data. However, their analysis differs from ours in the selection of stars, their spatial distribution, as well as in the derivation of distances to these stars, but nevertheless their resulting radial profile agrees remarkably well with our analysis, shown in the top panel of Fig.~\ref{fig:data_vr}. Based on the similarity of their results when splitting their data sample into the Northern and Southern Galactic hemisphere, they conclude that stellar streams or mergers are unlikely to cause this velocity pattern, but rather the gravitational pull of spiral arms. If that is indeed the case they deduce that the Outer Arm located around $R\approx 15$~kpc would have caused the observed transition in the radial velocity profile from positive to negative values, which is in agreement with our results. 

Several other studies have investigated stellar kinematics in order to deduce the spiral structure of the Milky Way disk. 
Using RAVE data \citet{Siebert2012} analyzed an observed gradient in the Galactocentric radial velocities in the immediate solar neighbourhood (within $2$~kpc distance from the Sun), which they argue is likely to arise from a non-axisymmetric potential perturbation due to spiral arms. Similar to our analysis they apply an analytic model of a long-lived spiral arm and conclude that their best model suggests a two-armed spiral perturbation, for which they estimate a density contrast of $14\%$ compared to the background density, which is in good agreement with our results. 

\citet{GrosbolCarraro2018} analyzed the barycentric radial velocities of $\approx1500$ A and B stars from the Sun towards the Galactic center in order to derive a model for the spiral potential of the Milky Way. They also conclude that their observed gradient in the radial velocities is best described by a fixed Galactic potential with an imposed two-armed spiral potential. However, they argue that the two major arms within the Milky Way should be the Perseus Arm around $R\sim 10$~kpc, as well as the Scutum Arm around $R\sim 5$~kpc, which differs from the predictions of stellar overdensities from our steady--state toy model. 

A similar gradient in radial velocities has been observed in the outer part of the disk between $10\,{\rm kpc}\leq R\leq15\,{\rm kpc}$ by \citet{Harris2019}, who investigated the kinematics of A and F stars within the disk along two pencil-beams sightlines. They observe small wiggles around $v_R=0$, which they attribute to noisy data due to their limited sample size. The authors conclude that their results do not show any clear evidence for spiral arms. 

The predicted overdensities from our steady-state toy model at the $R\approx 8$~kpc and $R\approx 15$~kpc spatially coincide with predictions from previous studies for the locations of the Local spiral Arm and the Outer Arm, respectively, which are shown in  Fig.~\ref{fig:comparison}. The inferred locations of spiral arms from other work are based on the surface density of neutral hydrogen gas \citep{Levine2006}, the distribution of interstellar dust \citep{Rezaei2018}, as well as the location of molecular masers \citep{Reid2019}. \citet{Miyachi2019} find an overdensity in red giant stars selected from \gaia\ and \zmass\ located close to the Local Arm. 

Our results agree with the predictions from \citet{Levine2006} for the Outer Arm, and with the predicted overdensities by \citet{Rezaei2018}, \citet{Reid2019} and \citet{Miyachi2019} at the location of the Local Arm. However, other studies suggest the presence of several additional features, which we do not observe in the Galactcocentric radial velocity profile. Most prominently, we do not see evidence for an overdensity at the location of the Perseus arm at $R\approx10-11$~kpc, which has been observed as a stellar overdensity \citep[e.g.][]{Monguio2015, Reid2019}, as well as in \ion{H}{1} gas \citep[e.g.][]{Levine2006}. 


Small offsets between the various predictions for overdensities and the location of spiral arms could arise due to different tracers. \citet{HouHan2015} find an offset between tangencies of spiral arms observed in gaseous tracers compared to tracers of old stars, although these displacements are small, i.e. $\sim1^{\circ}-5^{\circ}$, comparable to the expected width of the spiral arms. 

Thus such small offsets cannot explain the missing spiral arms that were observed by other authors. However, one potential explanation for our results is that the Local Arm as well as the Outer Arm are currently in a growing phase, whereas the Perseus Arm is decaying, which has been suggested previously by \citet{Baba2018}, who found evidence for a divergence in the stellar velocity field at the location of the Perseus Arm \citep[see also][]{Miyachi2019}. The predicted location of the Perseus Arm at $R\approx10$~kpc coincides with a blue--red gradient (with increasing $R$) in our Galactocentric radial velocity map, indicating an underdensity in the gravitational potential, and thus stars are on average moving away from this region. 

There are several other studies analyzing stellar overdensities within the Milky Way disk that have been interpreted as spiral structure. Almost two decades ago, \citet{DrimmelSpergel2001} already analyzed near- and far-infrared photometry from the COBE survey predominantly tracing red giant stars and cold dust, respectively, to study the spiral structure within the solar neighborhood. They found evidence for a two-armed spiral component, as well as a warped Galactic disk \citep[see also][]{Drimmel2000}. Subsequently, \citet{Benjamin2005} found several stellar enhancements in mid-infrared photometry that were associated with spiral arm tangencies \citep[see also][]{Benjamin2008}. 
Furthermore, an overdensity of red clump stars has been discovered based on near-infrared photometry from the VISTA Variables in the Via Lactea (VVV) ESO public survey extending from behind the Galactic bar, potentially tracing the spiral structure of the Perseus arm beyond the bulge \citep{Gonzalez2011, Gonzalez2018, Saito2020}.

\section{Summary and Discussion}\label{sec:summary} 

In this paper we present the detection of a non-axisymmetric spiral signature in the stellar kinematics in the Milky Way's mid--plane, observed in the mean Galactocentric radial velocities of luminous red giant stars. 
The observed pattern has a pitch angle that is comparable to pitch angles suggested in the literature for the Milky Way's spiral arms as shown in Fig.~\ref{fig:comparison}. Thus, we construct a simple toy model to interpret the observed signature, assuming that the feature arises due to a non-axisymmetric perturbation in the gravitational potential of our Galaxy caused by a two--armed logarithmic spiral rotating with a fixed pattern speed. 

Under these assumptions and a chosen pattern speed of $\Omega_{\mathrm{p}}=12\,\rm km\,s^{-1}\,kpc^{-1}$ ($\Omega_{\mathrm{p}}=2\,\rm km\,s^{-1}\,kpc^{-1}$) we estimate a maximum surface density of the perturbation at the solar radius of $\Sigma_{\rm max}(R_{\odot}) \approx 5.5\,M_{\odot}\,\rm pc^{-2}$ ($\Sigma_{\rm max}(R_{\odot}) \approx 6.0\,M_{\odot}\,\rm pc^{-2}$) and a pitch angle of the spiral signature of $p\approx0.21$ ($p\approx0.22$), i.e. $p\approx12.0^{\circ}$ ($p\approx12.7^{\circ}$), which is consistent with previous studies \citep[e.g.][]{Vallee2015}. We obtain an estimate for the scale length of the density perturbation that is large compared to the extent of the disk covered by our data, indicating that the amplitude of the perturbation stays approximately constant. While the statistical uncertainties on these model parameters are small, the uncertainties are dominated by systematic errors from the choice of model. Combined with previous studies of the local disk density, we find a surface mass density contrast of approximately $10\%$ at the solar radius with an increase towards larger radii. 

Note that the fundamental measurement we conduct is a radial velocity difference at different parts of the Galactic disk, which can be connected to a potential perturbation when assuming a steady state. What we would be most interested in of course would be to connect this potential perturbation to a total density perturbation, but this requires a global potential model and a disk-to-halo mass ratio. In our current model, we simply assume that the level of potential perturbation is roughly proportional to the level of density perturbation, i.e. we can obtain the density perturbation via Poisson's equation (see Eqn.~\ref{eq:poisson}). However, if the disk-to-halo mass ratio changes, this dependency would change as well, and thus the exact level of the density perturbation depends on the global potential model, as well as on how close the disk is to a steady state. 

Our model predicts overdensities arising from the non--axisymmetries in the gravitational potential, which approximately coincide with previous predictions for the locations of the Local Arm as well as the Outer Arm. However, we do not find any evidence for overdensities at the locations of other spiral features claimed in the literature, such as the Perseus Arm. These results could potentially be explained if the Local and Outer Arm are currently in a growing phase, whereas the Perseus Arm is decaying, which has been suggested by previous studies \citep{Baba2018, Miyachi2019}. 


Note that the perturbation in the gravitational potential that gives rise to the observed kinematic pattern in the Milky Way disk stars could also arise from resonances of the Galactic bar \citep[e.g.][]{Monari2016_bar} or a major merger of the Milky Way with a satellite galaxy in the past \citep[e.g.][]{Quillen2009}. In order to make further progress and potentially determine the origin of the observed kinematic signature, this signature has to be transformed to a global surface density map of the Milky Way's disk. This is most likely more complex than suggested by our simple toy model due to the unknown origin of the spiral pattern, the transient nature of spiral arms \citep[e.g.][]{Carlberg1985, DeSimone2004, Sellwood2014, Hunt2018}, as well as the stringent model assumptions of a steady state and a disk-to-halo mass ratio that are mentioned above. 

In order to understand the kinematics \textit{as well as} the overall density of the disk stars, and to construct a mapping of the total stellar surface density to the kinematic signature, a well understood selection function of the \apogee\ and \gaia\ surveys would be required. While several studies have determined the \apogee\ selection function \citep[e.g.][]{Bovy2014, Frankel2019}, the \gaia\ selection function remains only very approximately known to date. A well-understood selection function for the stars would permit direct measurement of the stellar density perturbations associated with these kinematic spiral signatures. 
Future comparison of observations like these to simulations of Milky Way analogues \citep[e.g.][]{Buck2020} with spiral arms, galactic bars, and/or an active merger history will shed new light on the origin of the observed kinematic spiral pattern. 


\acknowledgements
The authors would like to thank Ortwin Gerhard, Larry Widrow, Sarah Pearson, Cara Battersby, Sara Rezaei Khoshbakht, and Francesca Fragkoudi for interesting discussions and helpful feedback. Furthermore, we would like to thank the anonymous referee for their thorough and constructive comments, which significantly improved our manuscript. 

This project was developed in part at the 2019 Santa Barbara Gaia Sprint, hosted by the Kavli Institute for Theoretical Physics at the University of California, Santa Barbara. 

ACE acknowledges support by NASA through the NASA Hubble Fellowship grant $\#$HF2-51434 awarded  by  the  Space  Telescope  Science  Institute,  which  is  operated  by  the   Association  of  Universities for  Research  in  Astronomy,  Inc.,  for  NASA,  under  contract  NAS5-26555.

This work was supported by the Deutsche Forschungsgemeinschaft (DFG, German Research Foundation) -- Project-ID 138713538 -- SFB 881 (``The Milky Way System'', subproject A03). 

NF acknowledges support from the International Max Planck Research School for Astronomy and Cosmic Physics at the University of Heidelberg (IMPRS-HD).

JASH is supported by a Flatiron Research Fellowship at the Flatiron institute, which is supported by the Simons Foundation. 

JBF acknowledges support by the grant Segal ANR-19-CE31-0017 of the French Agence Nationale de la Recherche.

TB acknowledges support by the European Research Council under ERC-CoG grant CRAGSMAN-646955. 

\appendix 
\section{Vertical velocity structure of the Milky Way disk}

\begin{figure}[!t]
\includegraphics[width=\textwidth]{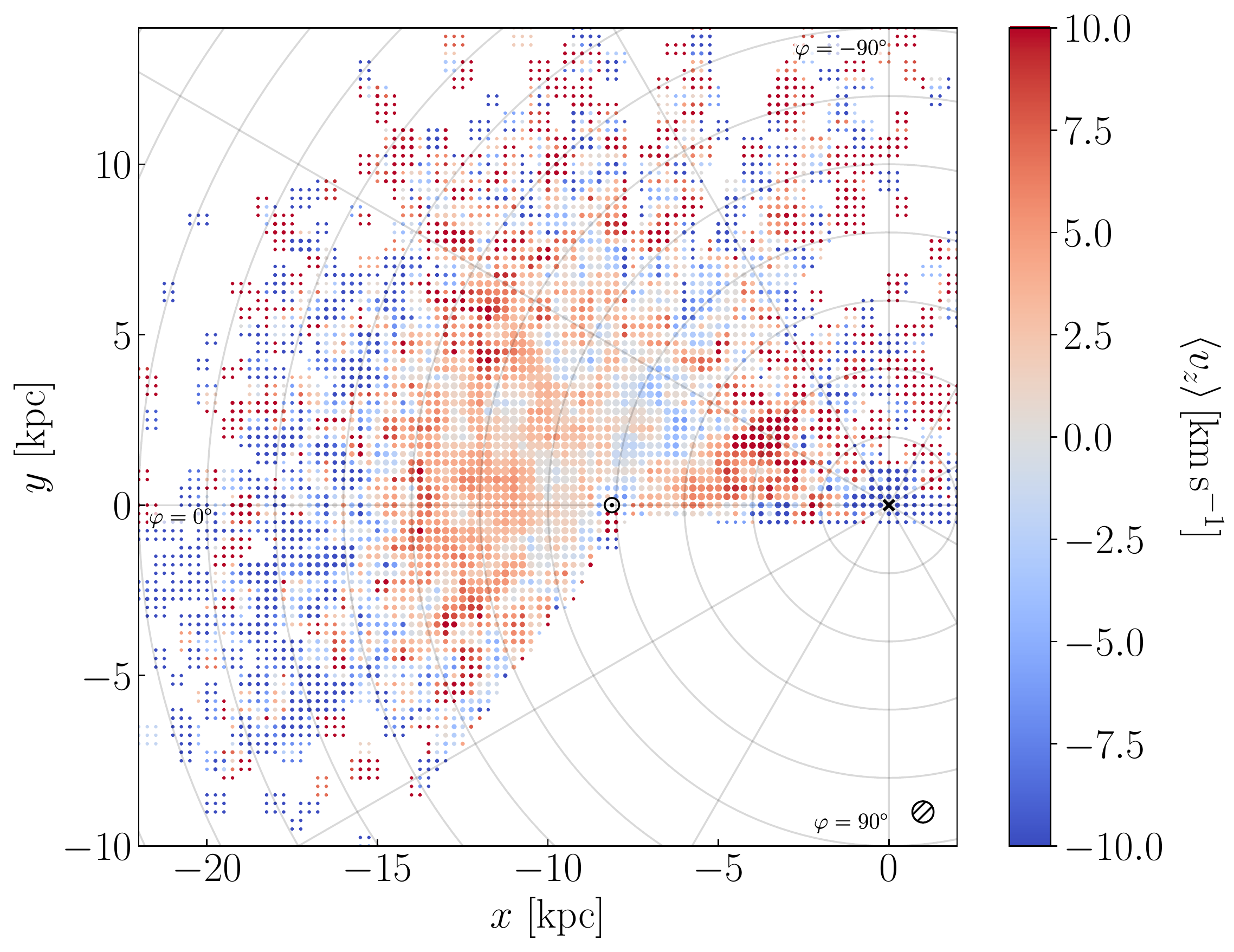}
\caption{Same as Fig.~\ref{fig:data} but colored by the mean vertical velocity component $\langle v_z\rangle$. \label{fig:vz}}
\end{figure}

In Fig.~\ref{fig:vz} we show a map of the average vertical velocity $\langle v_z\rangle$ of the stars in our sample. The same selection criteria and cuts have been applied as for Fig.~\ref{fig:data}. We would expect the vertical out-of-the-plane motion to decouple from the radial in-plane motion of the stars in a thin disk \citep[see \S~3.2 in][]{BinneyTremaine}. Nevertheless the map of $\langle v_z\rangle$ shows a qualitatively similar feature in the outer disk as observed in the Galactocentric radial velocities $\langle v_{R}\rangle$, although more ``smeared out'', i.e. the decrease in velocities around $R\approx 8-10$~kpc, which we observed in $\langle v_R\rangle$ is less pronounced in $\langle v_z\rangle$. 

Our results agree with Fig.~$3$ panels (C) and (D) in \citet{Poggio2018}, who analyzed the vertical velocities $v_z$ obtained from astrometric data from \gaia\ DR2 of upper main sequence and giant stars within $7$~kpc distance from the Sun. They observe a gradient in $v_z$ of $5-6\,\rm km\,s^{-1}$ between Galactocentric radii of $8$ to $14$~kpc, which spatially coincides with the increase in $\langle v_R\rangle$ from our observations. The authors interpret the observed feature as a signature of the Galactic warp. However, our data of the vertical velocity beyond $R\gtrsim15$~kpc reveal a decrease in $\langle v_z\rangle$ at larger radii, which resembles the pattern observed in $\langle v_R\rangle$, and therefore the observed pattern in $\langle v_z\rangle$ could potentially also be a consequence of spiral arms. 

While it would clearly be interesting to further investigate the $3$-dimensional motions of Milky Way disk stars and their potential correlation, this analysis is beyond the scope of the paper. 

\bibliography{literature_mw}

\end{document}